\documentclass{aastex62}
\begin{document}

\title{Toward the Detection of Relativistic Image Doubling in Imaging Atmospheric Cerenkov Telescopes}
\author[0000-0002-0786-7307]{Robert J.~Nemiroff}
\email{nemiroff@mtu.edu}
\affiliation{Michigan Technological University\\1400 Townsend Drive\\Houghton, MI 49931,USA}
\author{Neerav Kaushal}
\email{kaushal@mtu.edu}
\affiliation{Michigan Technological University\\1400 Townsend Drive\\Houghton, MI 49931,USA}
 
\begin{abstract}
Cosmic gamma-ray photons incident on the upper atmosphere create air showers that move to the Earth's surface with superluminal speed, relative to the air. Even though many of these air showers remain superluminal all along their trajectories, the shower's velocity component toward a single Imaging Atmospheric Cherenkov Telescope (IACT) may drop from superluminal to subluminal. When this happens, an IACT that is able to resolve the air shower both in time and angle should be able to document an unusual optical effect known as relativistic image doubling (RID). The logic of RID is that the shower appears to precede its own Cherenkov radiation when its speed component toward the IACT is superluminal, but appears to trail its own Cherenkov radiation when its speed component toward the IACT is subluminal. The result is that the IACT will see the shower start not at the top of the atmosphere but in the middle -- at the point along the shower's path where its radial velocity component drops to subluminal. Images of the shower would then be seen by the IACT to go both up and down {\it simultaneously}. A simple simulation demonstrating this effect is presented. Clear identification of RID would confirm in the atmosphere a novel optical imaging effect caused not by lenses but solely by relativistic kinematics, and may aid in the accuracy of path and speed reconstructions of the relativistic air shower.   
\end{abstract}


\section{Introduction} \label{sec:intro}

Air showers may result when a high energy particle enters the Earth's atmosphere. One such particle -- hereafter assumed to be a gamma ray for simplicity -- interacts high up in the atmosphere and starts a cascade of secondary particles, many of which move faster than the local atmospheric speed of light. These superluminal charged secondaries trigger the isotropic emission of Cherenkov light \citep{1958NCim....7..858M} by the atmosphere immediately in their wake. 

Practically, the gamma-ray must have an energy of at least a few hundred GeV to produce Cherenkov light visible on the ground \citep{1977SAOSR.381.....P}. For a primary photon at one TeV energy, about 100 photons per m$^2$ are seen on the ground to a detection radius of about 125 meters \citep{1977SAOSR.381.....P}. The Cherenkov photons arrive within a very short time interval, a few nanoseconds \citep{2006APh....25..391H}. Images obtained with Imaging Atmospheric Cherenkov Telescopes (IACTs) show the projected track of the air shower, which points back to the celestial object where the incident gamma ray originated. Past and current IACTs including Whipple \citep{1990ExA.....1..213L}, HEGRA \citep{1997APh.....8....1D}, VERITAS \citep{2002APh....17..221W}, MAGIC \citep{2004NIMPA.518..188B}, and H.E.S.S. \citep{2006A&A...457..899A}, have been instrumental in discovering and studying many astronomical sources of high energy radiations. A prominent future IACT is planned by CTA \citep{2011ExA....32..193A}.

Recent advances in understanding the appearance of objects moving faster than light through a given medium has revealed an unusual optical phenomenon known as ``relativistic image doubling" (RID) \citep{1971Sci...173..525C, 2015PASA...32....1N, 2018AnP...53000333N, 2019AAS...23325101N}. Specifically, after the component of the speed of a superluminal object towards an observer (radial velocity) drops from superluminal to subluminal, an observer, here an IACT, can perceive two images of that object simultaneously -- with one image moving forward along the expected track, and a second image moving backwards along the original track. RID effects have recently been hypothesized to help explain light curve features in gamma-ray bursts \citep{2019ApJ...inpress}. 

Although an example of RID has been recovered in the laboratory \citep{2016SciA....2E1691C}, its occurrence has not been clearly isolated, as yet, elsewhere. Previously, though, papers about IACT imaging have briefly indicated that such an effect could occur. For example, \citet{1982JPhG....8.1475H} mentions that ``At about 100-150 m from the axis the radiation from several heights arrives simultaneously". The effect was also touched on in \citet{1999APh....11..363H} with the statement "In particular, the profiles in the 60 m to 120 m distance range show a parabolic rather than linear shape, with photons from both the head and the tail end of the shower arriving late." This parabolic shape was evident in a few of the curves in their predictive Figure 9a. Furthermore, \citet{1999APh....11..363H} indicate an actual detection by a HEGRA IACT in their Figure 13b, although the plotted error bars appear to obscure its statistical significance.

More recently, \citet{2009APh....30..293A} reported on Monte Carlo simulations in support of the MAGIC IACTs that show ``In case of a small impact parameter (IP $\le$ 60 m), the light emitted in the higher part of the shower (the shower head) will arrive delayed with respect to the light emitted in the lower part of the shower (the tail)" and ``Events with an intermediate impact parameter show a flat time profile." Many of the simulated ovals in Figure 2 of \citet{2009APh....30..293A} indicate that the central part of some IACT images from MAGIC are expected to arrive about a nanosecond before the outer parts of the image ovals. 

The record of the Cherenkov light from an atmospheric shower by an individual telescope is typically a fleeting oval of illuminated pixels \citep{2006A&A...457..899A}. Although these images are usually published without indication of time-resolution beyond the likely duration of the event, which is a few nanoseconds, sub-event timing at a resolution of two-nanoseconds -- or below -- can be recorded \citep{2006APh....25..391H, 2017ICRC...35..747M}.

Although RID effects are inferred by these references, very few details are given. In particular, how and why two images of the shower are created together, how they diverge, how they are related to the Cherenkov cone, and the speeds and brightness of the individual images have been left undiscussed. 

Visually, a single observer or IACT will see Cherenkov emission at its brightest when the cluster of particles in the descending air shower is at the specific angle corresponding to the edge of the Cherenkov cone \citep{1963NucIM..20..263B}. As will be made clear, RID effects {\it create} the Cherenkov cone, making these effects the rule for air shower detections, rather than the exception. The double-image creation event may be obscured, however, near the creation point by air-shower width, so that some air showers will show diverging images more clearly than others. 

It is the purpose of this manuscript to conceptually isolate and study RID effects visible to IACTs, highlighting the temporal and angular scales that are most prominent. Section \ref{sec:RIDC} will focus on the concepts behind RID, while \ref{sec:RIDM} will present useful formulae. Subsequently, Section \ref{sec:RIDE} will present the results of a simple example simulation, while Section \ref{sec:Discussion} concludes with some discussion. 


\section{Relativistic Image Doubling: Concept} \label{sec:RIDC}

Assume that an air shower descends vertically toward the ground at a constant speed $c_{air} < v < c$, where $c$ is the speed of light in vacuum, $v$ is the speed of the air shower, and $c_{air}$ is the speed of light in air. Further assume that the shower is observed by a single IACT, modelled as a single point located at a distance $L$ from the base of the shower. Consider that $v$ and $c_{air}$ are constants and so not a function of the shower height $z$. These simple approximations are chosen to demonstrate RID effects without conflating it with other atmospheric and geometric effects. 

In the example scenario described, a cosmic ray photon strikes the atmosphere at height $z = h$ at $t = 0$. Initially, the angular height of the shower from the IACT vertex is $\theta = \arctan (h/L)$. The height of the shower at time $t$ is therefore $z = h - v t$. The component of the shower's speed directly toward the IACT is $v_r = v \sin \theta$, where $v_r$ is the shower's radial component toward the IACT. 

A critical angle occurs when $v_r(z) = c_{air}$, where the IACT is at an angle $\sin \theta_C = c_{air}/v$ with respect to the path of the shower. This angle arises when the shower is at a critical height $z_C$, quantified below in Eq.~\ref{z_C}. Near this shower height, the speed of the shower toward the IACT closely matches the speed of the Cherenkov light toward the IACT. Therefore a relatively large amount of this light arrives at the IACT "bunched up" -- over a very short time. At height $z_C$, it can be said that the observer is on the "Cherenkov Cone". Note that since Cherenkov light is emitted uniformly and isotropically along the shower trajectory, the Cherenkov Cone is an observer-dependent phenomenon. 

For most photons striking the Earth's atmosphere, an IACT will never be inside the shower's Cherenkov Cone for any value of $z$. Therefore, for these showers, the radial speed of the shower toward the IACT is never greater than $c_{air}$. These distant IACTs will see the shower go through the atmosphere, from top to ground, monotonically. This occurs for all IACTs far away from the shower, specifically when $L > h \cot \theta$.

Conversely, for IACTs close to the shower base, where $L < h \cot \theta$, will be inside the Cherenkov Cone for at least some shower heights $z$. These IACTs can see a RID event. Specifically, there will be a region starting at the top of the atmosphere and extending down to $z_C$ where $v_r > c_{air}$. In this region, the speed of the shower toward the IACT is {\it faster} than the speed of the Cherenkov radiation it triggers. When eventually seen by the IACT, this part of the shower will be seen time-backwards, meaning that Cherenkov radiation emitted in shower locations increasingly earlier will reach the IACT increasingly later. This part of the shower will appear to go {\it up} from $z_C$. After the shower has descended down past $z_C$ however, the radial speed of the shower will be {\it slower} than its Cherenkov light and so this part of the air shower will appear normally: time forwards and headed down.

The first Cherenkov light detected by the IACT will be emitted when the shower is exactly at the height $z = z_C$, when the IACT is on the shower's Cherenkov Cone, so that $v_r = c_{air}$. After that. the IACT receives Cherenkov light from two locations: from both above and below $z_C$, {\it at the same time}. This is the basis for relativistic image doubling (RID) \citep{2017arXiv170305811N, 2018AnP...53000333N}.

The RID image-pair creation event is the brightest part of perceived Cherenkov emission -- subsequently each of the diverging images will appear to fade as it moves away \citep{2018AnP...53000333N}. In essence, it is claimed here that some atmospheric Cherenkov showers, if resolved in both time and space, could appear to start at a point towards the middle of its path and then appear to a single IACT to move both up and down simultaneously.

\section{Relativistic Image Doubling: Math} \label{sec:RIDM}

We now derive equations and algorithms relevant to demonstrating how RID appears in air showers in the example scenario. The location of the actual shower at time $t$ is not the location of the two {\it images} of the shower visible to the IACT at time $t$, because it takes a significant time for light to go from the shower to the IACT. At height $z$, the actual radial velocity of the shower is 
\begin{equation} \label{v_r}
    v_r(z) = { z v  \over \sqrt{L^2 + z^2} } , 
\end{equation}
with the shower's actual transverse velocity $v_t(z)$ determined by $v^2 = v_r^2 (z) + v_t^2 (z)$. To calculate the perceived positions and perceived speeds of the shower images, light travel time must be incorporated. Starting at the top of the atmosphere, the total time $t_{total} (z)$ for Cherenkov light to reach the observer is the addition of two parts: $t_{shower} (z)$, the time it takes for the shower to reach height $z$, and $t_{light} (z)$, the time it takes for Cherenkov light to go from height $z$ to the IACT. Using straightforward geometry it is clear that 
\begin{equation} 
    t_{shower} = (h - z)/v ,
\end{equation}
while 
\begin{equation}
    t_{light} = \sqrt{L^2 + z^2} / c_{air} ,
\end{equation}
so that  
\begin{equation} \label{ttot}
    t_{total} = t_{shower} + t_{light}  .
\end{equation}

The first height where the shower is seen at the IACT, $t_{min}$, occurs when $t_{total}$ is a minimum. The corresponding height $z_C$ can be found from when $d t_{total} / dz = 0$ such that  
\begin{equation} \label{z_C}
    z_C = { {c_{air} L} \over { \sqrt{ v^2 - c_{air}^2} } } .
\end{equation}
The value of $t_{min}$ can be found by substituting $z_C$ into Eq.~\ref{ttot}. At later times, two images of the shower are visible simultaneously that have heights both above and below $z_C$. These are found by solving Eq.~\ref{ttot} for $z$, yielding   
\begin{equation} \label{zimages}
    z_{\pm} = { { c_{air}^2 t_{total} v - c_{air}^2 h   \pm 
              \sqrt{c_{air}^2 h^2 v^2 - 2 c_{air}^2 h t_{total} v^3
              + c_{air}^2 L^2 v^2 + c_{air}^2 t_{total}^2 v^4 - L^2 v^4 } }
              \over
              {v^2 - c_{air}^2} }  ,
\end{equation}
where $z_+$ and $z_-$ are the height of the shower's images, with $z_+$ always being above $z_C$, and $z_-$ always being below. 

The speed of each image can be computed by taking $v_{\pm} = d z_{\pm} / dt_{total}$, and the angular speeds ${\dot \theta}_{\pm}$ can be computed by dividing the transverse component of each image speed by its distance from the IACT. To estimate the apparent brightness of each image, we start from the assumption that the shower is intrinsically uniform and isotropic everywhere along its path. If the apparent brightness was proportional the absolute brightness, it could be estimated by simply noting the uniformly bright path lengths visible to the observer over a uniform time interval. However, since relatively long path lengths are seen over a uniform time interval when $v_r \sim c_{air}$, the apparent brightness $b$ of each image is proportional to the angular speed of the image. Since image brightness also falls by the square of its distance from the IACT, then  
\begin{equation} \label{Eqb}
   b_{\pm}  \propto { { {\dot \theta_{\pm}}} \over 
           { L^2 + z_{\pm}^2 } } .
\end{equation}
The brightness formally diverges at the start, when $v_r = c_{air}$ which occurs at
$z_{\pm} = z_C$, but this formal divergence is mitigated in practice by the shower being of finite angular size.

\section{Relativistic Image Doubling: Example} \label{sec:RIDE}

Specific results from a simple example are now presented. Here the shower is assumed to begin at height $h = 25$ km and go straight down. The distance of the IACT from the projected floor of the shower is assumed to be $L = 100$ meters. The index of refraction of the air is taken to be $n_{air} = 1.00029$, with the speed of light in air then being $c_{air} = c / n_{air}$. The speed of the shower is assumed to be constant at $v = c$. From Eq.~\ref{z_C}, the value of $z_C$, the height $z$ where $v_r(z) = c_{air}$, is found to be $z_C = 4.15$ km. Note that this $z_C$ is relatively low in the atmosphere compared to the 25 km height where the shower first developed. 

\begin{figure}[!htb]
            \centering
            \hspace*{-1.0cm}
            \includegraphics[scale=0.4]{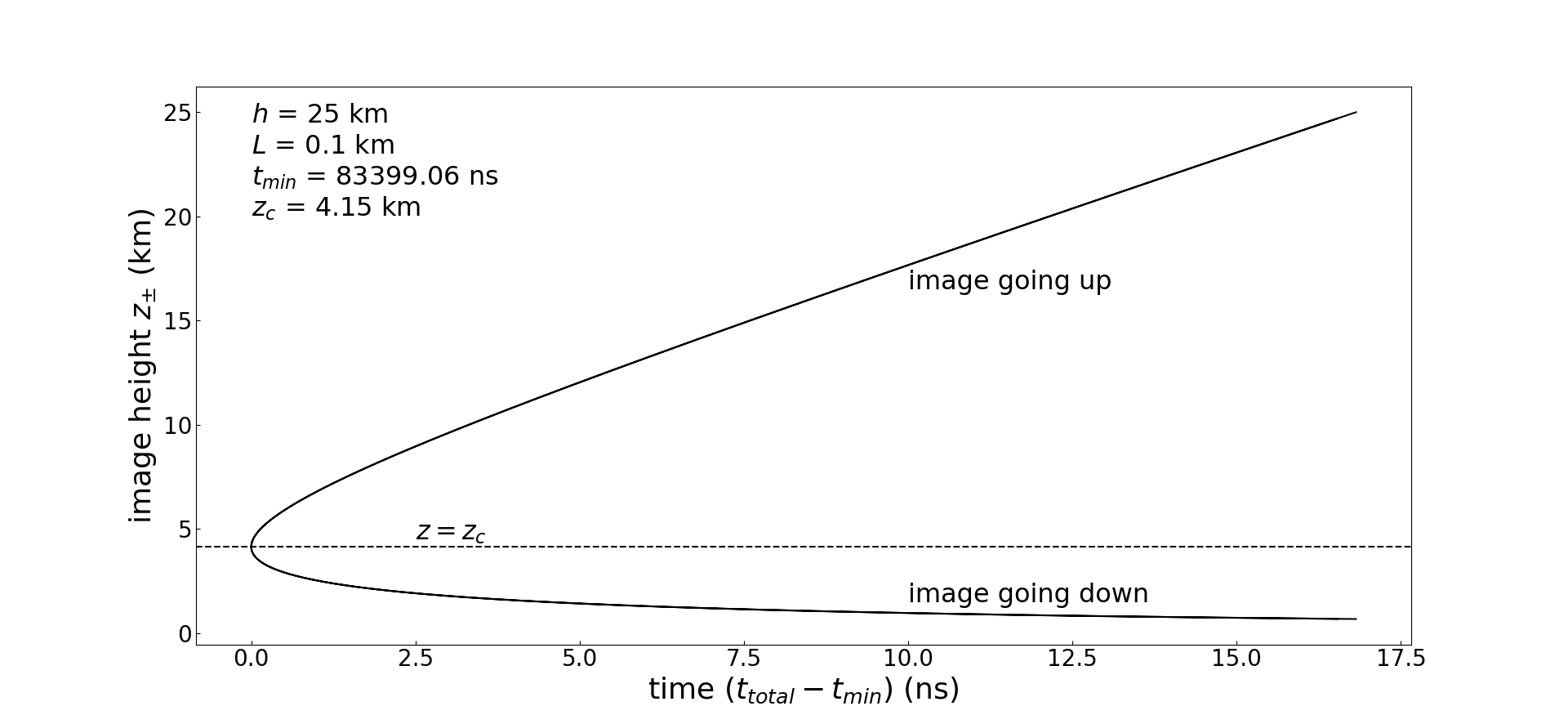}
        \caption{A plot of image height versus time. The RID image pair-creation event is first seen at the IACT when $(t_{total} - t_{min}) = 0$, where $t_{total}$ corresponds to the time between when the shower begins at $h = 25$ km, and the time when the shower is seen at the IACT. The time $t_{min}$ is the minimum of $t_{total}$ over the shower path. The IACT is located at $L = 100$ meters from the shower base. The shower is first seen at a height of $z_C = 4.15$ km. Afterwards, the image of the shower that appears to move upwards is shown by the upper line, while the image that appears to move downward is depicted by the lower line.}
            \label{fig:zt}
        \end{figure}

A plot of $t_{total}$ versus $z_{\pm}$, as computed from Eqs.~\ref{ttot} and \ref{zimages}, is shown in Figure~\ref{fig:zt}. Inspection of this figure clearly shows that $t_{total}$ is double valued with respect to $z$. The first time $t_{total}$ that the IACT sees the shower is when $z = z_C$, when RID first is seen.  

\begin{figure}[!htb]
            \centering
            \includegraphics[scale=0.4]{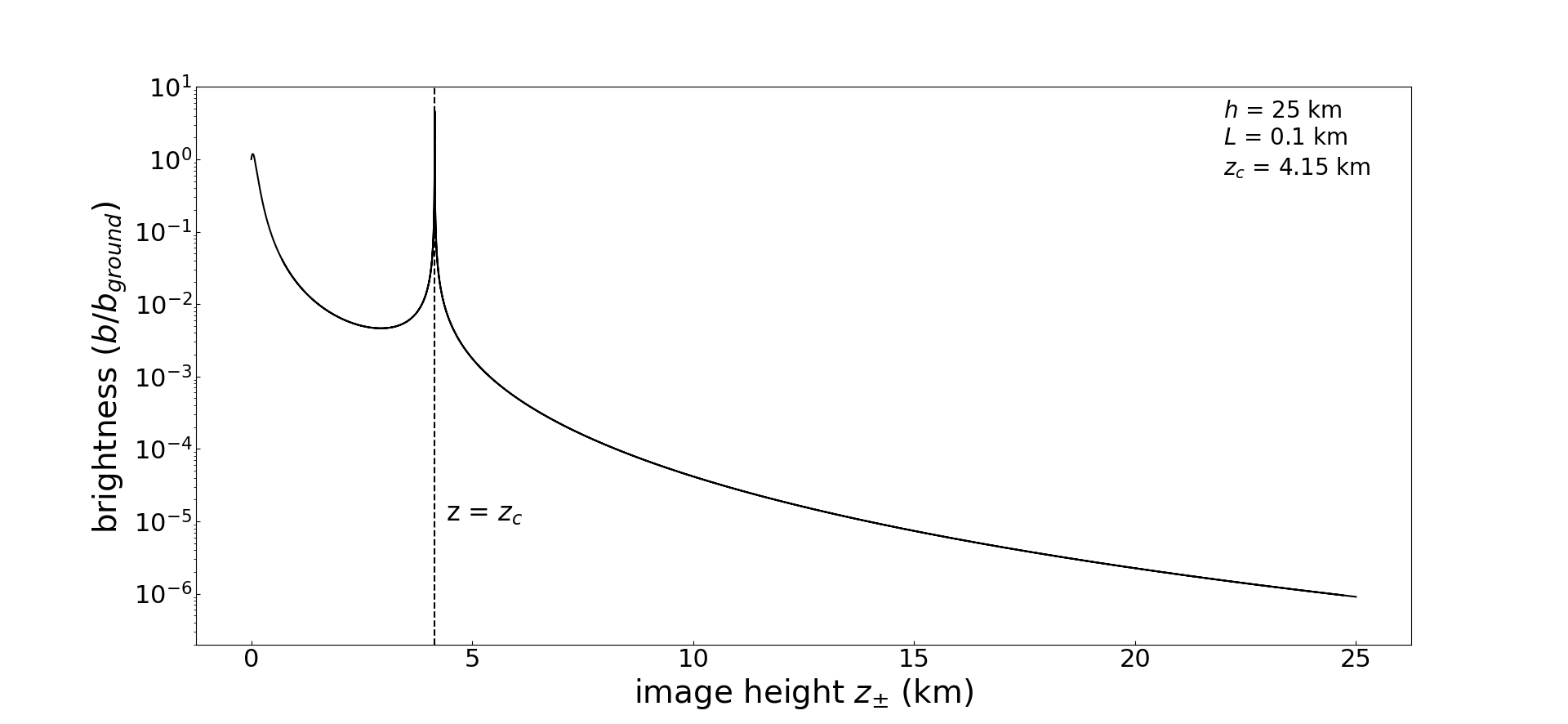}
        \caption{A plot of the apparent brightness of the shower images versus image height. The intrinsic brightness of the actual shower is assumed to be uniform and isotropic during the shower's decent. Apparent brightness $b$ is normalized to that seen just before the shower impacts the ground. The shower image that moves upward starts very bright at a height of $z_C = 4.15$ km, thereafter appearing continually dimmer to the IACT. Simultaneously, the shower image moving down from $z_C$ first drops quickly in brightness but then re-brightens when it nears the IACT. }
            \label{fig:bz}
        \end{figure}

A plot of the relative brightness of each image as a function of $z_{\pm}$, computed from Eq.~\ref{Eqb}, is shown in Figure~\ref{fig:bz}. This figure shows a sharp peak in the brightness of each image when $v_r = c_{air}$ at $z = z_C$. The lower image is also relatively bright when near the IACT. 

\begin{figure}[!htb]
            \centering
            \hspace*{-1.0cm}
            \includegraphics[scale=0.4]{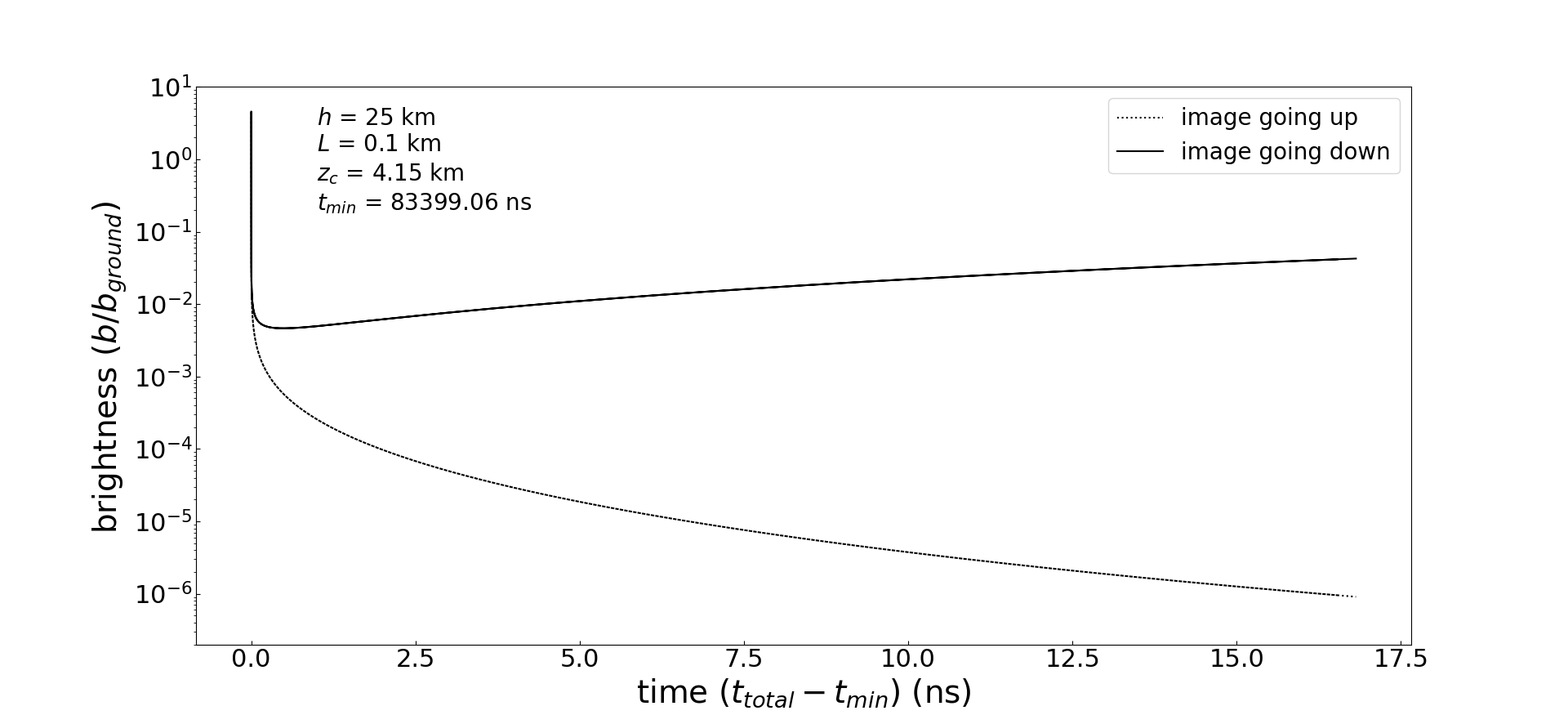}    
            \caption{A plot of apparent image brightness versus time. Both images are very bright at the RID image-pair creation event, when they are first seen at $z = z_C = 4.15$ km, but then fade rapidly. The image headed upward becomes significantly more faint because it becomes significantly farther from the IACT. }
            \label{fig:bt}
        \end{figure}

The apparent brightness of each image is shown as a function of time $t_{total}$ in Figure~\ref{fig:bt}. This figure shows the light curves that would be seen by the IACT if the IACT had negligible size. The figure shows that at first the IACT sees nothing, but then suddenly, the shower is seen at height $z_C = 4.15$ km, with a sharp spike in brightness. This corresponds to the image pair-creation episode of the RID, when the IACT is on the shower's Cherenkov Cone. As the images move away from $z_C$ -- one image moves up to toward the top of the atmosphere, while simultaneously the second image moves toward the ground. The top image disappears when it reaches the top of the atmosphere, while the bottom image disappears when it hits the ground. The time axis of Figures~\ref{fig:zt} and \ref{fig:bt} are truncated when the upper image exits the top of the atmosphere. 

\section{Discussion} \label{sec:Discussion}

Although the concept has been alluded to previously, few details about the appearance of RID in air showers have been published and therefore may be generally unknown. Some of these novel, interesting, and potentially useful details of are summarized here. 
First, the relative brightness of the two images that appear in descending air shower have never been noted previously. For the simple example case explored here, relative image brightness can be computed from Eq.~\ref{Eqb} and is evident in Figures \ref{fig:bz} and \ref{fig:bt}. In general, the two images in RID will appear to be created with equal apparent brightness but the perceived ratio of their brightnesses will quickly diverge. This attribute of air shower RID can be important for identifying and verifying images as resulting from RID. Furthermore, individually, both RID images are brightest just when they appear to be created, but the brightness of both images will appear to fade rapidly. In the simple example case studied, this fading is shown in Figures \ref{fig:bz} and \ref{fig:bt}.

The sudden appearance of the RID images in air showers is novel, without classical precedent, and has not been published previously. Counter-intuitively, there will be absolutely no flux from an air shower received by an IACT before it detects the shower's RID image creation event. It is not that the shower appears undetectably faint by the IACT at earlier times, but rather that no light at all from the descending shower will have reached the IACT before the arrival of the RID event. Then, suddenly, a bright RID event becomes visible. This is evident, for example, in Fig. \ref{fig:zt}.

The kinematic context of {\it why} air shower RID images are so bright has also never been discussed before in the detail described here. The underlying physical reason for the high brightness of the RID images is that the radial speed of the air shower toward the IACT drops from superluminal to subluminal. The closer the radial speed of the shower to the speed of light toward the IACT, the larger the segment of the air shower seen by the IACT in a given amount of time. This corresponds to the passing of the Cherenkov cone over the IACT, but the underlying kinematics should not be obscured by citing this association abstrusively. In fact, the Cherenkov cone is a simple observer effect created precisely by this velocity coincidence \citep{2018AnP...53000333N}. 

It has also not been previously noted that the RID images appear with formally infinite angular speed. After creation, the speed of each diverging image appears to drop rapidly. This behavior is evident from the discussion surrounding Eq.~\ref{Eqb}. This is important for detection and tracking of RID images by IACTs.

Another novel but previously-unnoted attribute of air shower RID images is that one image of the RID pair is seen moving time backwards -- back up into the upper atmosphere. This image moves up from the $z_C$ because it is being seen at the IACT at progressively earlier times. In the simple example explored here, this behavior is clear from Eq.~\ref{zimages} and Fig.~\ref{fig:zt}. However, the time-backwards nature of upward moving images is generic to all shower RIDs. This may be important, for example, because it shows that even though early parts of the air shower are over, they may not be gone from view. Information about this early part of the shower may still be recovered in the next few nanoseconds because the images above $z_C$ always arrive late. 

Another newly-noted feature of tracking individual images of an air shower RID is that the downward moving image (only) may pass relatively near the IACT as it approaches the ground -- and so may appear to re-brighten.  Therefore, if this image becomes too dim to track, it may subsequently become visible to this IACT for a second time. This second visibility may occur even though this image is {\it not} related to the Cherenkov cone of the descending shower. This behavior results solely from the decreasing $r$ in the $1/r^2$ brightness falloff, and is evident in the example in Eq.~\ref{zimages} and the left most part of Fig.~\ref{fig:bz}.

RID effects may be thought immeasurable, by some, because both images fade so rapidly. However, even tracking the images for a short angular distance would be interesting. It is unclear from a previous reference whether an RID effect has actually already been observed. Inspection of Figure 13b of the HEGRA paper \citet{1999APh....11..363H} does indicate such an effect, but the error bars are large and its statistical significance was not estimated -- it could well be marginal. Calling new attention to this effect and giving previously unpublished details may create a higher priority for existing and future IACTs to become equipped with sensitive and high speed imagers that could better detect RID and track the two diverging images.

One might argue that RID is a trivial effect. Who cares that two images of the same shower are sometimes visible simultaneously? In our view, however, RID is a novel optical effect caused not by lenses but solely by relativistic kinematics -- and so is interesting basic physics even without a demonstrated usefulness. However, discovering and tracking air shower RID effects may even prove useful. Such observations could add information that better allows shower orientation to be recovered, or give an independent method of confirming air shower geometry. In the simple case discussed above, the ground location of the air shower $L$ might be better located by use of Eq.~\ref{Eqb}, for example. 

Studying the basic science of other multiple-imaging optical effects is common practice in gravitational lensing \citep{1987PhDT........12N, 1992grle.book.....S} and temperature inversions effects in the Earth's atmosphere \citep{1999YoungsFamousWebPages}, for example. 

Along this line, it is interesting to further compare, conceptually, RID to gravitational lensing. Multiple imaging in gravitational lensing is created by similarities in the magnitudes of the time delay caused by the increased slowing of time near a gravitational mass, and the time delay caused by the increased path length further from the line of sight. These similarities creates different locations in the plane of the lens that have the same total time delay, including critical points on the time-delay surface where images form.

Similarly, RID in air showers is created by the similarities in magnitude of the time delay for the shower to reach given heights, and the time delay caused by the different path lengths from those heights to the IACT. This similarity creates different locations along the shower's path that have the same total time delay, including critical locations where images form.

Why have RID effects remained so obscure? One reason is that they have no classical analog -- they depend crucially on the finiteness of the speed of light. Classical thinking may allow scientists to accurately visualize how the air shower itself descends, but not always how it appears as it descends. 

Another reason for the sparsity of previous RID analyses is that RID events happen locally on time scales too short for humans to notice: nanoseconds as shown by Fig.~\ref{fig:zt}. However, as computer technology and miniaturization is increasing the frame rate that can be captured, imagers appearing over the past few years are becoming capable of recording sub-nanosecond events \citep{2016SciA....2E1691C}.

There are also attributes of realistic IACTs that obscure RID effects. One is that the IACT dish itself is not just a point -- it may be so large that light travel time across its combined mirrors is significant when compared to $t_{light}$ (B. Humensky, private communication, 2019). Then any light curve that the IACT measures will convolve the size and shape of the IACT, not just geometry inherent to the air shower. However, even in this case, if the IACT's angular point spread function is smaller than the angular trajectory of the shower on the IACT's imager, then the dish-crossing time will not compromise the general character of the shower's temporal development on the IACT's imager. In general, no matter the size and geometry of the IACT, the central angular point of the shower at $z = z_C$ will begin to light up first, after which two images of the shower will appeare to diverge. 

Another obscuring practical outcome is that RID effects will look different to different IACTs, even for the same air shower. The ground locations of individual IACTs are paramount -- only those close enough to the base of the air shower so that $z_C < h$ will have a Cherenkov cone pass over them and so be able to see an RID effect. For the example given, from Eq.~\ref{v_r} and $h=25$ km, any IACT with $L > 603$ meters will not see the shower start with a RID. Therefore, simply adding together the images of multiple IACTs will likely not enhance the detection of RID effects, and may even convolute it beyond recognition. However, a careful reconstruction accounting for the timing separate RID events as seen by different IACTs should be possible that could enhance RID detection.

There are many interesting RID-related effects that were deemed too complex to be incorporated in the primarily conceptual work. One such interesting case occurs when the bulk speed of the particles in the air shower decreases significantly along its path to the surface. This may happen even if the shower remains superluminal and continually triggers Cherenkov radiation. If $v$ decreases, then $v_r$ will decrease proportionately, forcing $z_C$ to occur higher in the atmosphere. Additionally, air density and ionization could cause significant variations in $c_{air}$. A full treatment of these effects would likely require more detailed and complex computer modeling.

In sum, resolving relativistic image doubling events in time and space would recover a novel type of optics caused not by lenses but by relativistic kinematics alone. Further, tracking simultaneous images as they emerge and diverge in angle, angular speed, and relative brightness could resolve or independently confirm
information about the air shower, including shower trajectory, speed, and brightness along its path as deconvolved through equations in Section 3. These may, in turn, result in an increased accuracy in the direction of the originally incident photon or cosmic ray that started the air shower. 

\section*{Acknowledgments} \label{sec:A}
We thank Oindabi Mukherjee and Rishi Babu for valuable comments.

\end{document}